\documentclass[12pt]{iopart}

\expandafter\let\csname equation*\endcsname\relax
\expandafter\let\csname endequation*\endcsname\relax
\usepackage{amsmath}
\usepackage{iopams}  
\usepackage{graphicx}
\usepackage{epstopdf}
\usepackage{amssymb}
\usepackage[colorlinks=true,urlcolor=blue,citecolor=blue,linkcolor=blue,breaklinks=true]{hyperref}
\usepackage{color}
\usepackage{ulem}
\usepackage[dvipsnames]{xcolor}

\begin{document}

\title[Strongly correlated superconductivity with long-range spatial fluctuations]{Strongly correlated superconductivity with long-range spatial fluctuations}

\author{Motoharu Kitatani$^1$, Ryotaro Arita$^{1,2}$, Thomas Sch\"afer$^3$,  Karsten Held$^4$}

\address{$^1$RIKEN Center for Emergent Matter Science (CEMS), Wako, Saitama, 351-0198, Japan \\
$^2$Department of Applied Physics, The University of Tokyo, Hongo, Tokyo, 113-8656, Japan\\
$^3$Max-Planck-Institute for Solid State Research, 70569 Stuttgart, Germany \\
$^4$Institute of Solid State Physics, TU Wien, 1040 Vienna, Austria}
\ead{kitatanimo@gmail.com}
\begin{indented}
\item \today
\end{indented}

\begin{abstract}
We review recent studies for superconductivity using diagrammatic extensions of dynamical mean field theory. These approaches take into account simultaneously both, the local correlation effect and spatial long-range fluctuations, which are essential to describe unconventional superconductivity in a quasi-two-dimensional plane. The results reproduce and predict the experimental phase diagrams of  strongly correlated system such as cuprates and nickelates. Further studies reveal that the dynamical screening effect of the pairing interaction vertex has dramatic consequences for the transition temperature and may even support exotic mechanisms like odd-frequency pairing. We also discuss the dimensionality of layered materials and how to interpret the numerical results in two dimensions.
\end{abstract}

\section{Introduction}
Layered and thin-film superconductivity has been a central topic in condensed matter physics due to its fascinating properties. Especially after the discovery of the cuprate  \cite{Bednorz1986} and iron-based \cite{Kamihara2008} superconductors, such quasi-two-dimensional systems can be considered central for unconventional superconductivity \cite{Keimer2015,Ishida2009}. In 2019 superconductivity of thin nickelate films was discovered \cite{Li2019}. For understanding these systems, it is essential to describe appropriately correlated electrons of $3d$-transition metals, including spatial fluctuation effects that are amplified by the quasi-low dimensionality.


For the theoretical understanding of unconventional superconductivity, Feynman diagrammatic perturbation techniques have often been used \cite{Yanase2003,Scalapino2012}. Such schemes can describe spatial fluctuation effects by collecting relevant diagrams, and they have succeeded in describing spin-fluctuation mediated superconductivity \cite{Scalapino1986,Scalapino1987,Bickers1989,Kuroki2008}. On the other hand, these methods have difficulties to capture non-perturbative phenomena such as the correlation-induced metal-insulator (Mott) transition \cite{Mott1968,Imada1998}, one of the cornerstones of correlated electron systems. In this regard, the dynamical mean field theory (DMFT) is the method of choice for treating Mott transitions \cite{Georges1992,Georges1996}. DMFT becomes exact in the limit of infinite spatial dimensions \cite{Metzner1989}. In contrast, spatial fluctuation effects ignored by DMFT become more relevant for low dimensional systems, e.g., layered materials or thin films that are further away from this limit.

Against this background there have been many attempts to capture both the strong correlation effects and (long-ranged) spatial fluctuation effects simultaneously. In particular, two main types of extensions of DMFT for including such spatial fluctuations have been established: one class are the cluster extensions which use the numerically exact self-energy a self-consistently determined cluster model \cite{Hettler1998,Lichtenstein2000,Kotliar2001,Maier2005}. While it can describe the overall structure of the phase diagram of unconventional superconductors like cuprates \cite{Maier2000,Gull2013}, it has been gradually understood that the finite size effect of the cluster is severe, e.g. for capturing the gap structure induced by strong spin fluctuation in the intermediate coupling regime \cite{Schaefer2015,Simkovic2020,Schaefer2021}. The other class of extensions is the diagrammatic route \cite{Rohringer2018}. As ordinary diagrammatic methods, we need to choose how to collect diagrams beyond the DMFT starting point and a systematic improvement is cumbersome \cite{Ribic2017}.
Considering additional physical requirements such as causality \cite{Backes2020arXiv} is thus helpful. On the other hand, one can address more exotic situations including lower temperatures, longer ranged correlations and multi-orbital \cite{Galler2016} physics. Thus these approaches can, e.g., capture accurately the (pseudog-)gap opening due to the strong long-ranged magnetic fluctuations \cite{Schaefer2015,Schaefer2021,Kitatani2020,Klett2021}. One can analyze even extremely fluctuating systems like (quantum) critical regimes \cite{Rohringer2011,Antipov2014,Schaefer2017,Schaefer2019} and estimate the degree of anisotropy of electronic correlations in layered materials \cite{Klebel2021}.

In this article, we review recent studies of unconventional superconductivity with diagrammatic extensions of DMFT. This article is organized as follows. In Sec.~2 we overview methods for normal state calculations and the analysis of the superconductivity. In Sec.~3 we argue some relevant (but rarely discussed) points for studying the layered and thin-film superconductivity within these frameworks. In Sec.~4 we show recent results on the superconductivity. We finally note the summary and outlooks in Sec.~5.

\section{Overview of the method}
In this section, we provide an overview on methods called "{\it diagrammatic extensions of DMFT}" (GW+(E)DMFT, FLEX+DMFT, dynamical vertex approximation (D$\Gamma$A), dual fermion, TRILEX, DMF$^2$RG,...) \cite{Sun2002,Biermann2003,Hague2004,Sadovskii2005,Kusunose2006,Toschi2007,Rubtsov2008,Held2008,Katanin2009,Rohringer2013,Taranto2014,Gukelberger2015,Kitatani2015,Valli2015,Ayral2015,Li2015,Ayral2016,Tomczak2017}. First, we explain the essence of these schemes, and then show how to treat superconductivity. Here, we focus on the generic part of these methods. For the detailed explanations and their comparison, please see the review article~\cite{Rohringer2018}.

\subsection{Normal state calculations}
\begin{figure}[t!]
        \centering
                \includegraphics[width=1.0\linewidth,angle=0]{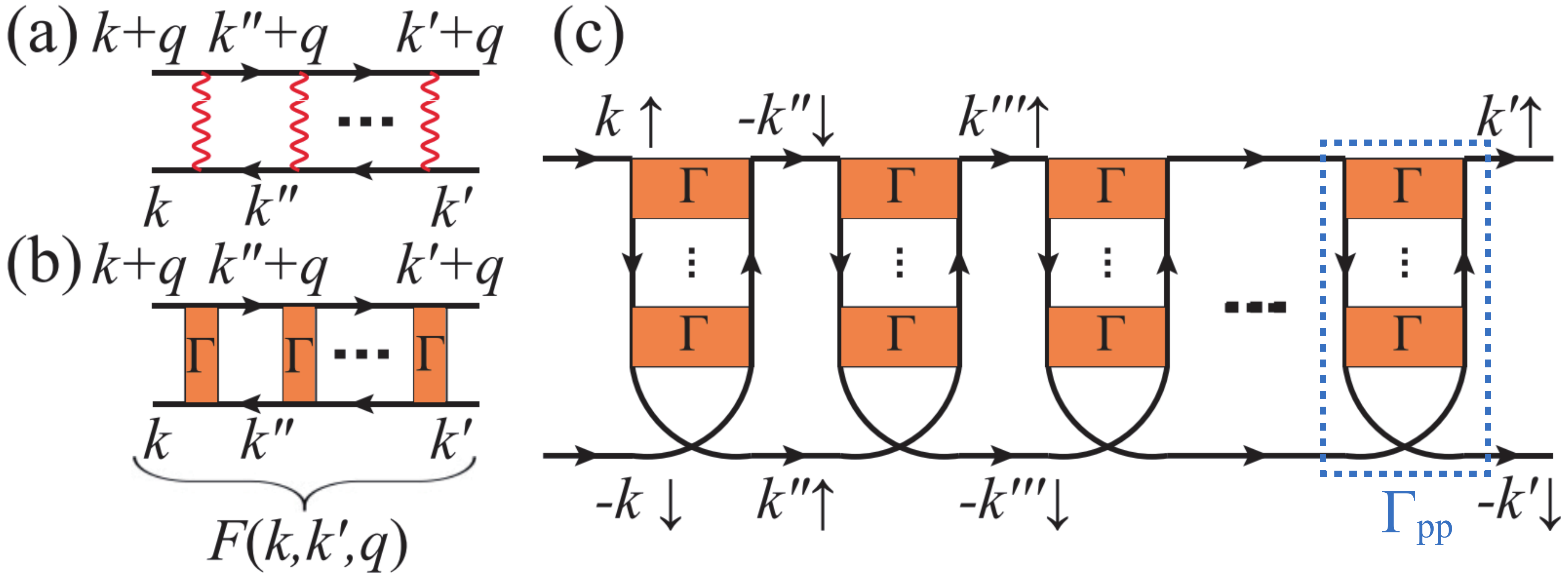}
        \caption{(a) Particle-hole ladder diagrams (solid line: the Green function) representing antiferromagnetic spin fluctuations for interaction $U$ (red wiggly line). (b) More general particle-hole diagrams, including vertex corrections in strongly correlated systems ($\Gamma$: the irreducible vertex for the particle-hole channel). (c) Such spin fluctuations act as a pairing glue for superconductivity in the particle-particle channel. Here, an exemplary diagram for the particle-particle irreducible vertex $\Gamma_{\rm pp}$ (blue dotted box) is shown where $\Gamma_{\rm pp}$ is made up from the particle-hole fluctuations from (b). Taken from Ref.~\cite{Kitatani2019}.
        }
        \label{fig:schematic}
\end{figure}

Since unconventional superconductivity is regularly found in the vicinity of a magnetically ordered state \cite{Keimer2015,Ishida2009}, it is widely believed that spin fluctuation effects should be an (if not {\it the}) important key for understanding unconventional superconductivity. Often these systems are modeled by a single-band Hubbard model \cite{Hubbard1963,Hubbard1964,Kanamori1963,Qin2022,Arovas2021}. Also for this model, fluctuation diagnostic analyses could show the importance of spin fluctuations in the pseudogap regime \cite{Gunnarsson2015,Wu2016,Schaefer2021b}. Diagrammatically a particle-hole ladder diagram is relevant for spin fluctuations. We show the simplest example in Fig.~\ref{fig:schematic}(a), where the ladder is mediated by the bare interaction (the on-site repulsion $U$ for the simple Hubbard model). Such diagrams are used, e.g., in the fluctuation exchange (FLEX) approximation \cite{Bickers1989}. When we consider the relevant diagrams from a DMFT starting point (for instance in the dynamical vertex approximation \cite{Toschi2007,Katanin2009}) similar particle-hole diagrams enter, but now mediated by the irreducible vertex $\Gamma$ instead of $U$. Here, an irreducible vertex is a diagram that can not be split into two pieces by cutting two internal Green function lines in the corresponding channel: particle-hole (ph) or particle-particle (pp). This quantity can be derived from the one- and two-particle Green functions $G,\chi_{\rm ph}$ through the inverse Bethe-Salpeter equation: 
\begin{equation}
\Gamma^{\omega=0}_{\rm ph} = (\chi^{\omega=0}_{\rm ph})^{-1} - (\chi^{\omega=0}_{0,{\rm ph}})^{-1},
\end{equation}
where $\chi_0$ is the bare (bubble) susceptibility. Please see Refs.~\cite{Rohringer2012,Wentzell2020} for detailed explanations for the vertex function itself. Then, the diagram for the fully reducible vertex $F$ becomes like Fig,~\ref{fig:schematic}(b). From it we can calculate the self-energy by using the equation of motion (EOM, also known as Schwinger-Dyson equation in this context) which is an exact relation connecting the vertex and the self-energy. The basic concept is common to all diagrammatic extensions while the building block differs: the bubble (and ladder) diagrams in GW+DMFT (FLEX+DMFT) \cite{Sun2002,Biermann2003,Gukelberger2015,Tomczak2017,Kitatani2015}, the fully reducible local vertex in dual fermion \cite{Rubtsov2008} or DMF$^2$RG \cite{Taranto2014}, $\Gamma$ in ladder D$\Gamma$A \cite{Toschi2007,Katanin2009} and the fully irreducible vertex in parquet D$\Gamma$A \cite{Valli2015}, the three point fermion-boson vertex in TRILEX \cite{Ayral2015} etc.. In ladder schemes which we regularly use nowadays, we need to choose one relevant channel. Since the Coulomb interaction is repulsive and we often study the repulsive Hubbard model, we usually pick up particle-hole ladder diagrams. If we study the attractive model, the relevant channel will change and we should first pick up particle-particle diagrams \cite{DelRe2019}. Furthermore, solving the parquet extension (treating the contributions of all channels on equal footing) from DMFT result have been developed \cite{Valli2015,Li2016} and recently applied to study superconductivity instability, however, for high temperatures \cite{Kauch2019arXiv}.

\subsection{Superconductivity analysis}
The general way for approaching the symmetry broken superconductivity state is the extension to Nambu space, which is quite hard to treat. Instead, we often linearize anomalous components focusing on the infinitesimal gap function using the normal state quantities only, which is enough for calculating the transition temperature. In this case, we need to solve the eigenvalue problem (linearized gap equation) for the superconducting (particle-particle) channel \cite{Otsuki2014,Kitatani2015},
\begin{equation}
    \lambda \Delta(k)
    =
    -\frac{1}{\beta N_k} \sum_{k^{\prime}} 
    \Gamma_{\rm pp}(k,k^{\prime},q=0) G(k^{\prime}) G(-k^{\prime}) \Delta(k^{\prime}),
    \label{eq:Eliashberg}
\end{equation}
where $G$ ($\Delta$) is the normal (anomalous) Green function, and $\Gamma_{\rm pp}$ is the irreducible vertex in the particle-particle (cooperon) channel, depending on the four-vectors $k, k^{\prime},q$ (momentum and Matsubara frequency), $\beta$ is the inverse temperature, $N_k$ is the number of k-points, and $\lambda$ is the eigenvalue of the kernel of the superconducting channel which is the key quantity. When $\lambda$ reaches unity the gap function has a finite solution, indicating the superconducting phase transition. In other words, $\lambda \rightarrow 1$ means the divergence of the Bethe-Salpeter equation in the particle-particle channel mediated by the pairing vertex $\Gamma_{\rm pp}$. A typical diagrammatic structure of the spin-fluctuation mediated $\Gamma_{\rm pp}$ is shown in Fig.~\ref{fig:schematic}(c). The vertex $\Gamma_{\rm pp}$ describes the scattering of Cooper pairs which corresponds to the pairing glue for superconductivity. As indicated in Fig.~\ref{fig:schematic}(c) in ladder D$\Gamma$A calculations we use the particle-hole ladder diagrams [blue dashed box or Fig.~\ref{fig:schematic}(b)] that correspond to spin (and charge) fluctuations to calculate $\Gamma_{\rm pp}$. In a more complete but also numerically much more expensive parquet calculation  \cite{Kauch2019arXiv}, also the feedback of the particle-particle fluctuations ofn the particle-hole fluctuations is taken into account.

One crucial issue is how to perform low-temperature calculations. This is always a problem for studying unconventional superconductivity because of its typically low transition temperature: $T_{\rm c}\!\lessapprox\!({\rm bandwidth})/500$. Since vertices depend on three frequencies, we can store and calculate only a limited number of Matsubara frequencies which is usually insufficient for studying unconventional superconductivity. We explain how to technically overcome this problem and summarize recent progress on this topic in the Appendix.

\section{Remarks on calculating superconductivity}
\subsection{Eigenvalues with divergence of $\Gamma$}
\begin{figure}[t!]
        \centering
                \includegraphics[width=1.0\linewidth,angle=0]{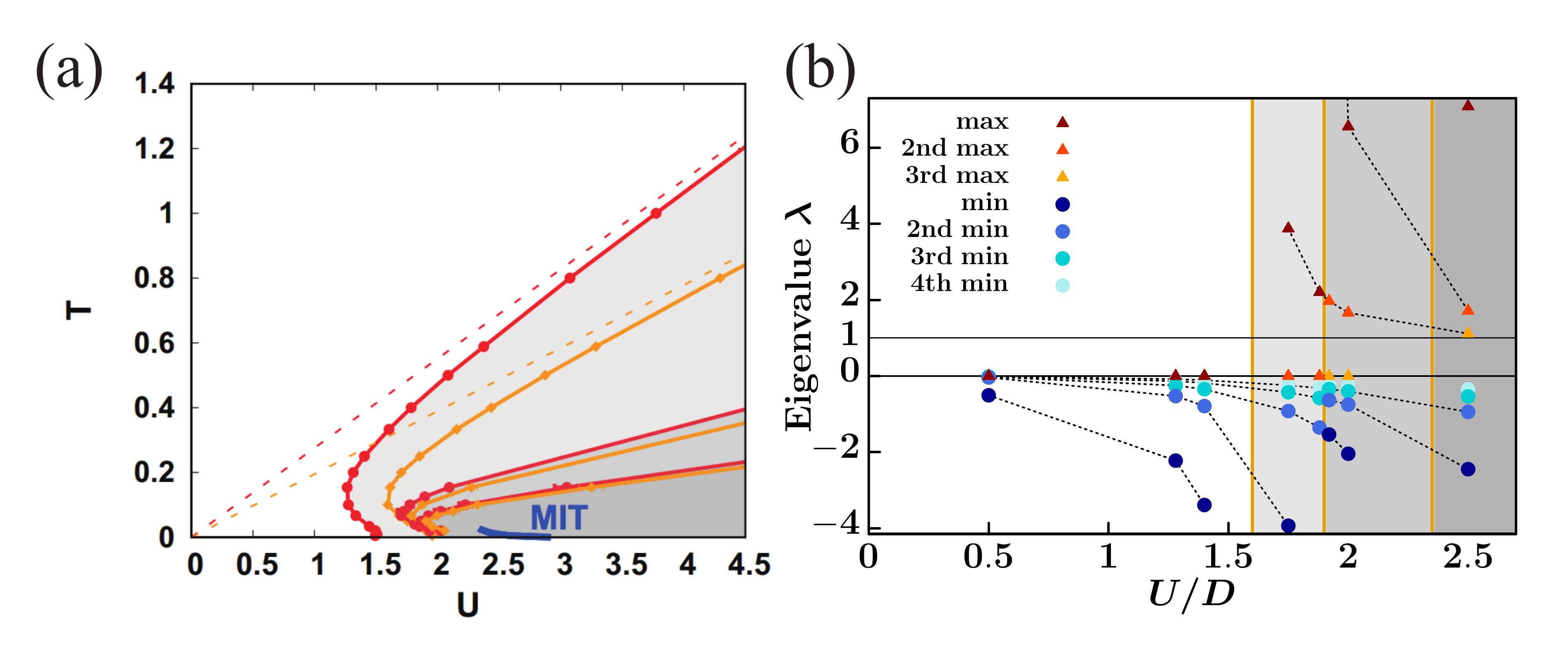}
        \caption{(a) DMFT $T$ vs.~$U$ diagram of the divergencies of the irreducible vertex for the square lattice Hubbard model without frustrations at half-filling (taken from Ref.~\cite{Schaefer2016}). The blue line indicates the Mott metal-insulator transition (MIT). The orange lines show the divergence of $\Gamma^{\nu,\nu^{\prime}(\omega=0)}_{\rm pp}$ at low frequencies (similarly, red lines are for the divergence of the irreducible vertex in the charge channel). (b) Leading eigenvalues (both maximum and minimum) of the local Eliashberg equation from the DMFT result at $T=0.1$. Vertical lines indicate the interaction values where the irreducible vertex diverges (orange lines in (a)) at this temperature. The energy unit is half the bandwidth ($D=4t$) -- also in (a).}
        \label{fig:divergence_gamma}
\end{figure}

As mentioned in the previous section, the fully reducible vertex function is calculated through the Bethe-Salpeter equation. The divergence of the Bethe-Salpeter series indicates a physical phase transition since it should connect to the divergence of the susceptibility in the corresponding fluctuation channel. However, it was found that the {\it irreducible vertex itself may diverge} without phase transitions \cite{Schaefer2013,Schaefer2016,Chalupa2018,Thunstrom2018}. In Fig.~\ref{fig:divergence_gamma}(a) we show red/orange lines where $\Gamma_{\rm d/pp}$ diverges without any physical instabilities in the DMFT calculation for the two-dimensional square lattice Hubbard model (taken from Ref.~\cite{Schaefer2016}). Please note that this divergence is not an artifact of the DMFT approximation and are already present in the (exactly solvable) Hubbard atom \cite{Thunstrom2018}. Also, some recent studies show that these lines change if we include spatial fluctuations \cite{Vucicevic2018}. It has been demonstrated recently that these divergences are responsible for a reduction in the local charge response \cite{Gunnarsson2017} and, at the same time, a strong enhancement of the charge compressibility in the vicinity of the Mott transition \cite{Reitner2020}.

Here, we additionally analyze the eigenvalue $\lambda$ of the Eliashberg equation when such divergence occurs. The leading eigenvalue in the superconductivity channel $\lambda$ is often referred to as a  ``measure of $T_{\rm c}$", since $\lambda$ usually monotonically increases as we decrease the temperature and $\lambda\!=\!1$ corresponds to the divergence of the superconducting susceptibility. However, this is not the case if divergences in the irreducible vertex occur, as shown in the following.

Let us consider eigenvalues in the local system as a simple example where the DMFT becomes exact for a corresponding Anderson impurity model. 
We then calculate the local Eliashberg equation
\begin{equation}
    \lambda \Delta(\omega_n)
    =
    -\frac{1}{\beta} \sum_{\omega_n^{\prime}} 
    \Gamma^{\rm loc,s}_{\rm pp}(\omega_n,\omega_n^{\prime},\omega_m=0) 
    G^{\rm loc}_{\rm DMFT}(\omega_n^{\prime}) G^{\rm loc}_{\rm DMFT}(-\omega_n^{\prime}) \Delta(\omega_n^{\prime}),
    \label{eq:local-Eliashberg}
\end{equation}
where the local Green function ($G^{\rm loc}$) depends on the Matsubara frequency $\omega_n$, and the local particle-particle irreducible vertex for the singlet channel ($\Gamma^{\rm loc,s}_{\rm pp}$) can be calculated within DMFT.

In Fig.~\ref{fig:divergence_gamma}(b), we show the (three maximal and four minimal) eigenvalues $\lambda$ of the local Eliashberg equation with the same model parameters as Fig.~\ref{fig:divergence_gamma}(a) at $T/D=0.1$, where $D=4t=1$ is the half-bandwidth. At first glance, the $U$-dependence of the eigenvalues is not systematic (dark red/blue points). For the region without any divergence in $\Gamma_{\rm pp}$ ($0<U/D<1.6$), all eigenvalues are smaller than zero, reflecting the fact that the repulsive interaction does not induce a momentum-independent $s$-wave superconductivity. However, as we cross the orange line at $U_1/D\!\approx\!1.6$ in Fig.~\ref{fig:divergence_gamma}(a) indicating the divergence in $\Gamma_{\rm pp}$, 
one large positive eigenvalue appears in the superconducting channel. When crossing the second divergence line at $U_2/D\!\approx\!1.9$, we can see that another large eigenvalue appears and a large negative eigenvalue disappears. Concomitantly, the eigenvalue $\lambda$ that was largest in $U_1<U<U_2$  smoothly connects to the second largest within $U_2<U<U_3$ ($U_3$: third divergence point), with the new large eigenvalue becoming the largest $\lambda$. Similarly, for the negative eigenvalues, the second lowest $\lambda$ within $U_1<U<U_2$  continues as the lowest $\lambda$  within $U_2<U<U_3$. The same happens again when crossing the third divergence line. This indicates that the large positive eigenvalue can appear through the infinitely large eigenvalue instead of crossing unity. The number of eigenvalues greater than unity corresponds to the number of vertex divergence lines crossed, but there is {\it no phase transition} as $\lambda$ remains $\neq 1$. 
This result can be naturally understood in terms of the divergence in $\Gamma_{\rm pp}$. Actually, the original article \cite{Schaefer2013} analyzed the eigenvalues of the generalized susceptibility $\chi$ and found that its eigenvalues can be negative. This corresponds to a divergence in the irreducible vertex and, hence, is represented by the orange lines in Fig.~\ref{fig:divergence_gamma}(b) if we think about $\chi \sim \frac{\chi_0}{1+\Gamma_{\rm pp} \chi_0}$.

In the following we show how that the superconducting phase transition in the strongly correlated regime is actually governed by the eigenvalues $\lambda$ which are close to unity, rather than by the leading (i.e. largest) eigenvalue.
Since the pairing interaction matrix $\Gamma^{\rm singlet,triplet}_{\rm pp}(k,k^{\prime})$ is Hermitian, we can perform an eigenvalue decomposition of the vertex. Taking care of the fact that the right and left eigenvectors of the Eliashberg equation (Eqs.(\ref{eq:Eliashberg},\ref{eq:local-Eliashberg})) are different, we can reconstruct the (irreducible and full) vertex by using the eigenvalues and (right) eigenvectors of the Eliashberg equation as
\begin{align}
    -\frac{1}{\beta N_k}
    G^*(k)\Gamma_{\rm pp}(k,k^{\prime},q=0)G(k^{\prime})
    &=\sum_{i} \lambda_i \Delta^*_i(k)G^*(k)G(k^{\prime}) \Delta_i(k^{\prime}), 
    \label{eq:lowrankGamma} \\
    -\frac{1}{\beta N_k}
    G^*(k)F_{\rm pp}(k,k^{\prime},q=0)G(k^{\prime})
    &=\sum_{i} \frac{\lambda_i \Delta^*_i(k)G^*(k)G(k^{\prime}) \Delta_i(k^{\prime})}{1-\lambda_i}. \label{eq:lowrankF}
\end{align}
Here, the right eigenvectors of the Eliashberg equation $\{\Delta_i\}$ are normalized as $\sum_k \Delta^{*}_i(k)G^*(k)G(k)\Delta(k)_j=\delta_{i,j}$ where $\delta_{i,j}$ is the Kronecker delta (which means the right and left eigenvectors make a bi-orthogonal basis set). Such a low rank decomposition of the vertex function is useful for approximating the complicated vertex structure, and indeed had been applied in the FLEX+T-matrix approach \cite{Yanase2001,Yanase2004} (while they treated as if $\Delta_i$ are orthogonal). 

Eqs.~(\ref{eq:lowrankGamma},\ref{eq:lowrankF}) clearly indicate that the divergence of $\lambda_i$ only connects to the divergence of $\Gamma$ 
[Eq.~(\ref{eq:lowrankGamma})]  and not to the divergence of fully reducible vertex $F$ 
[Eq.~(\ref{eq:lowrankF})] which is immediately relevant for physical phase transitions (since it contains all the vertex corrections of the physical susceptibility). Therefore, for the purpose of analyzing phase transitions, we {\it should not study the leading eigenvalue but eigenvalues close to unity} since $\lambda\rightarrow 1$ still means a diverging $F$ and, thus, a phase transition. One important open question is how the eigenvalues above unity behave, which would, according to Eq.~(\ref{eq:lowrankF}), be related to a negative contribution to the local susceptibility \cite{Gunnarsson2017}. 

\subsection{Estimation of the superconducting transition temperature from the purely two-dimensional system}\label{sec:MW}
For purely two-dimensional systems the Mermin-Wagner theorem \cite{Mermin1966,Hohenberg1967,Koma1992,Su1998} forbids continuous symmetry breaking at finite temperatures. Nevertheless, we sometimes refer to finite superconducting transition temperatures $T_{\rm c}$ [spontaneous continuous U(1)-gauge symmetry breaking]. To what extent is this justified?

\begin{figure}[t!]
        \centering
                \includegraphics[width=1.0\linewidth,angle=0]{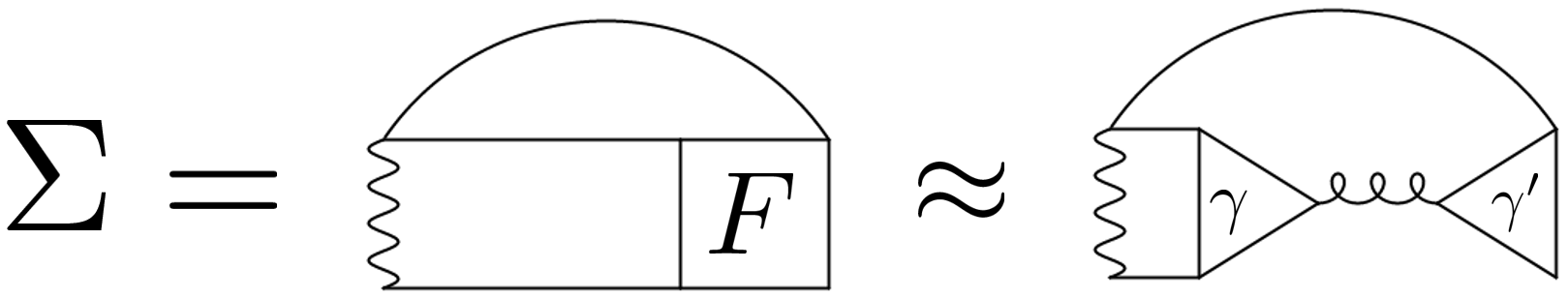}
        \caption{Diagrammatic representation of the equation of motion which connects between the self-energy ($\Sigma$) and the fully reducible vertex ($F$), the Green function (solid lines) and the bare interaction (wavy lines), which can be further approximated if we focus on a specific fluctuation (curly line).}
        \label{fig:EOM}
\end{figure}

First, let us explain how the Mermin-Wagner theorem can be understood (not proven) within diagrammatic calculations from the paramagnetic region. The self-energy $\Sigma(k)$ is connected to the fully reducible vertex function $F(k,k^{\prime},q)$ through the equation of motion, and, for the extremely fluctuating region, we can assume a specific channel dominates in the vertex contributions $F$ (Fig.~\ref{fig:EOM}) like,

\begin{align}
    \Sigma(k) &=
    -\frac{U}{\beta^2}
    \sum_{k^{\prime},q}
    F(k,k^{\prime},q)G(k^{\prime})G(k^{\prime}+q)G(k+q), \label{eq:EOM} \\
    &\approx-\frac{U}{\beta^2}
    \sum_{k^{\prime},q}
    \gamma(k,q)\chi(q)\gamma(k^{\prime},q)G(k^{\prime})G(k^{\prime}+q)G(k+q) \label{eq:EOM2} \\
    &\stackrel{\xi \rightarrow \infty}{\longrightarrow} 
    \int_{q \approx Q} \frac{{\rm d} q}{(q-Q)^2} f(k) \label{eq:MW-integration} \\
    &\longrightarrow  \left\{ \begin{array}{l}
     \;{\rm Power}\;{\rm law}\;{\rm divergence} \;\;({\rm in}\; {\rm 1D}),                \\
             \;{\rm Logarithmic}\;{\rm divergence} \;\;({\rm in}\; {\rm 2D}), \\
             \;{\rm Finite} \;\;({\rm in}\; {\rm 3D}),
    \end{array}\label{eq:12D} 
    \right.
\end{align}
where $\chi$ is the susceptibility of the relevant channel and $\gamma$ is the three-point vertex in that channel (c.f., \cite{Kontani2006,Vilk1997,Rohringer2016}). 
For the transformation from Eq.~(\ref{eq:EOM2}) to Eq.~(\ref{eq:MW-integration}), we assume an extremely fluctuating situation around $q \approx Q$ where the relevant susceptibility can be approximated as $\chi (q) \propto \frac{1}{(q-Q)^2+\xi^{-2}}$, where $\xi$ is the correlation length. 
Please note that for $\xi \rightarrow \infty$ we can replace $q$ by  $Q$ in all terms 
but $\chi(q)$, perform the $k'$ summation which yields a non-diverging term $f(k)$.
Now the crucial difference regarding dimension is that the self-energy diverges in a two-dimensional system while it converges to a finite value in three dimensions (for detail differences of the spin-fluctuation mediated self-energy between in 2D and 3D, see, e.g., \cite{Vilk1997,Rohringer2016}). Please note that this integral ($\int_{q \approx 0} dq/q^2$) is the key quantity for the rigorous proof \cite{Mermin1966} using Bogoliubov's inequality. Also, such a difference behavior for the long-range fluctuation is essential for determining the critical behavior \cite{Chubukov1994,Sachdev1995,Dare1996,Schaefer2019}. 

The difference in Eq.~(\ref{eq:12D}) indicates that, for a two-dimensional system, the self-energy must diverge as the susceptibility approaches infinity while it has not to be the case for a three-dimensional system. As the relevant susceptibility increases, the self-energy damping effect will win at some point, and then the susceptibility is not enhanced anymore (For the specific method, e.g., fluctuation exchange approximation, there is a more detailed analytical explanation of why FLEX satisfies the Mermin-Wagner theorem for antiferromagnetic long-range orders \cite{Kontani2006}). Also, for the $d$-wave superconductivity, it has been discussed that the similar singular structure in 2D suppresses $T_{\rm c}$ for the superconductivity long-range order \cite{Yanase2004}. In this regard, we can understand that we now obtain a finite superconducting $T_{\rm c}$ if we do not include the $d$-wave superconducting fluctuation effect to the self-energy.

For performing more appropriate calculations, we need to include all fluctuation effects and study the (quasi-two-dimensional) three-dimensional model, which is too complicated for analyzing electronic lattice models. This question has been addressed more carefully for spin systems (c.f., a study even for the electron lattice model if limited to spin fluctuation effects \cite{Dare1996}). In Ref.~\cite{Yasuda2005}, the authors performed the (quantum) Monte-Carlo simulations for (quasi-one/two-dimensional) three-dimensional systems. For both cases the  N\'{e}el temperature should go to zero as $J_{\perp}/J \rightarrow 0$, satisfying the Mermin-Wagner theorem ($J$ and $J_{\perp}$ are the in-plane and inter-plane Heisenberg coupling constant, respectively). On the other hand, they show quite different behaviors: $T_{\rm N} \propto -1/\ln{(\alpha J_{\perp}/J)}$ ($\alpha$: constant) in 2D while $T_{\rm N} \propto J_{\perp}/J$ in 1D probably reflecting the difference divergence strength in Eqs.~(\ref{eq:12D}). 

Indeed, the authors of Ref.~\cite{Yasuda2005} found a universal behavior of the transition temperature $T^{\rm AF}_{\rm c}$ in the weak inter-layer coupling region, indicating the relation between $T^{\rm AF}_{\rm c}$ in quasi-1D/2D system and the singular behavior in the purely 1D/2D system. In this region, the $T^{\rm AF}_{\rm c}$ is quite stable in 2D, which changes only $10-20\%$ if we change the coupling strength orders of magnitude $0.001<J_{\perp}/J<0.1$, which we can assume as the typical quasi-two-dimensional materials. In this sense, even without the inter-layer coupling, we can roughly define the transition temperature as the stable value of weakly coupled systems. We should also mention that such a very weak inter-layer coupling is not relevant at all if we do not include extremely fluctuating critical $d$-wave pairing fluctuations, exemplified by the fluctuation exchange calculations \cite{Arita2000}.

The question remains of how much is the effect of introducing the $d$-wave superconducting fluctuation effect to the self-energy in quasi-two-dimensional systems (i.e., without diverging effect on the self-energy described above).  While this point has not been fully understood yet, a dynamical cluster approximation (DCA, a cluster extension of DMFT) study \cite{Gunnarsson2015} showed that the $d$-wave superconducting fluctuation has only a tiny effect on the self-energy even close to the superconducting transition temperature (please note that due to the finite size effect of the clusters, this argument can not be applied to the ideal two-dimensional physics described above, while the calculation was done in pure two dimension). 

Summarizing the above-mentioned points, it is naively expected that without including the inter-layer coupling explicitly, we can roughly define the transition temperature in quasi-2D systems, which can be estimated by a purely two-dimensional system without $d$-wave pairing fluctuation effect on the self-energy (since the inter-layer coupling is too small to change the physics without critical $d$-wave pairing fluctuation \cite{Arita2000}, and for $d$-wave pairing fluctuation, the feedback effect for the self-energy is found to be minor \cite{Gunnarsson2015} except for singular behavior of long-range fluctuation limit in a pure two-dimensional system). This point would be in sharp contrast to quasi-one-dimensional systems where it will be quite difficult to perform the quantitative argument without explicitly defining the inter-layer coupling. 

Finally, $d$-wave pairing fluctuations would dominate if we approach $T_{\rm c}$ due to its divergent contribution in the purely two-dimensional model. In this sense, we can regard the current result as the extrapolation from a bit higher temperature calculations, where such critical fluctuations (introducing a singularity) are not relevant. The details of such a dimensional crossover on the superconductivity remain open. Further studies similar to Ref.~\cite{Yasuda2005} for the electronic lattice system are desired, i.e., including the $d$-wave superconducting fluctuation and comparing between purely two-dimensional systems and weakly coupled quasi-two-dimensional layered systems. We also note a possibility of the Berezinskii-Kosterlitz-Thouless (BKT) transition \cite{Kosterlitz1973} regarding the superconductivity in the two-dimensional Hubbard model \cite{Maier2005prl,Vilardi2020}. In this case, $T_{\rm c}$ remains as a quasi-long-range order even in  purely two dimensions, so that the quasi-two-dimensional $T_{\rm c}$ is rather stable against the inter-layer coupling. 

\section{Superconductivity properties}
In this section, we review recent studies for unconventional supercondcutivity by using  diagrammatic extensions of DMFT \cite{Otsuki2014,Kitatani2015,Otsuki2015,Kitatani2017,Jaksa2017,Kitatani2019,Vilardi2019,Sayyad2020,Kitatani2020,Grigory2020,Sayyad2021arXiv}. 

\begin{figure}[t!]
        \centering
                \includegraphics[width=1.0\linewidth,angle=0]{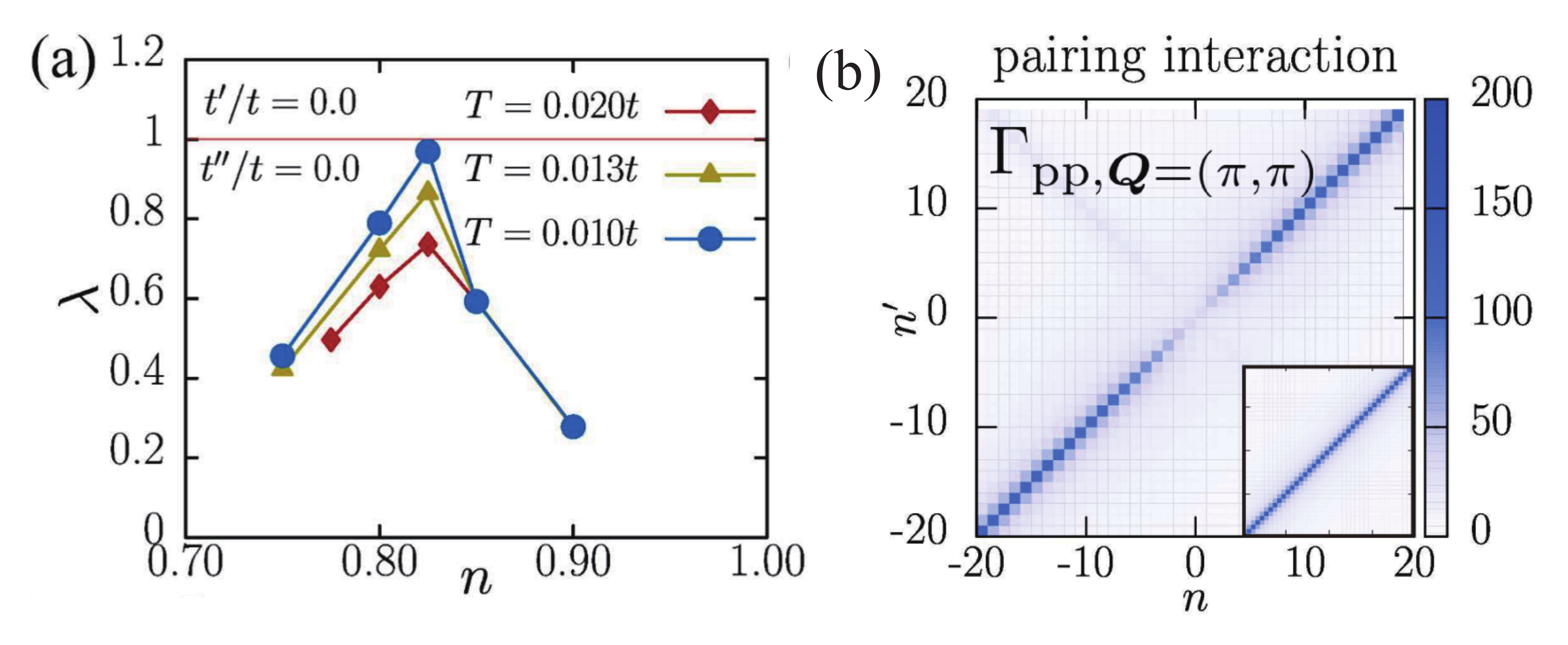}
        \caption{(a) The leading superconducting eigenvalue $\lambda$ against  filling $n$ in the unfrustrated square lattice Hubbard model with $U/t=6,\beta t =50,75,100$. (b) Dynamical vertex structure (Matsubara frequency dependence) of the particle-particle reducible (pairing) vertex, $\Gamma_{{\rm pp},Q=(\pi,\pi)}(\nu_n,\nu_{n^{\prime}},\omega=0)$ which includes spatial (in particular spin) fluctuations by means of D$\Gamma$A. We also show a typical structure of $\Gamma_{\rm pp}$ of mean-field-like approaches in the inset of (b). Taken from Ref.~\cite{Kitatani2019}}
        \label{fig:phase_pairingvertex}
\end{figure}

First, we start with the discussion of the phase diagram. For cuprates, the parent materials are antiferromagnetic charge-transfer insulators (often modelled as Mott insulators in a single-band model). As we dope carriers, superconductivity appears in some doping regime and then, upon further doping, the system eventually becomes a normal Fermi liquid. Many diagrammatic extensions of DMFT have succeeded in describing this behavior since they contain the spatial fluctuation beyond the local approximation. One crucial point is that these schemes can capture the critical divergent behavior of the spin susceptibility: $\chi \propto \exp({-\Delta/T})$ \cite{Katanin2009,Otsuki2014}, like two-dimensioal antiferromagnets \cite{Chakravarty1988}. 
For capturing such a critical behavior, the effect of long-range fluctuation is necessary. 
Properly evaluating such a strong spin fluctuation is essential for describing the strongly correlated superconductivity. It can be a glue for the superconductivity on one side, while at the same time, it can reduce the electron spectrum (and then $T_{\rm c}$) dramatically on the other side. Another critical point is the dome structure of $T_{\rm c}$. In Fig.~\ref{fig:phase_pairingvertex}(a), we show the D$\Gamma$A result for the leading eigenvalue in the two-dimensional square lattice Hubbard model without frustration, which exhibits the dome structure \cite{Kitatani2019}. While it is widely observed in cuprate materials, this dome structure was difficult to be captured by the FLEX-like diagrammatic approaches. Since we now consider the diagrammatic expansion starting from the DMFT, which takes the local strongly correlated effect, many methods have succeeded in capturing the dome structure of $T_{\rm c}$ \cite{Kitatani2015,Kitatani2017,Jaksa2017,Kitatani2019,Sayyad2020,Grigory2020,Sayyad2021arXiv}. Some FLEX+DMFT studies also discuss the relation between the Pomeranchuk (nematic) instability and superconductivity \cite{Kitatani2017,Sayyad2021arXiv}. The TRILEX study showed that the charge fluctuation becomes more relevant to the superconductivity for the triangular lattice models \cite{Cao2018}.

One remarkable feature of the vertex correction is the dynamical screening effect on the pairing interaction \cite{Kitatani2019,Grigory2020}. In Fig.~\ref{fig:phase_pairingvertex}(b), we show the pairing interaction mediated by spin fluctuations at $Q=(\pi,\pi)$, which depends on two fermionic Matsubara frequencies (with fixed bosonic frequency: $\omega=0$). We first see the strong intensity of the diagonal structure which corresponds the spin fluctuation mediated pairing: $\chi_{{\rm sp},Q=(\pi,\pi)}(\nu-\nu^{\prime})$. Additionally, we can see the strong screening effect in the low-frequency regime. This structure cannot be captured by the ordinary mean-field like (paramagnon mediated) interaction (as shown in the inset of figure~\ref{fig:phase_pairingvertex}(b)) since paramagnons depend on one bosonic frequency only.

This structure is general, and we observed it for all parameters of Fig.~\ref{fig:phase_pairingvertex}(a); see also Ref.~\cite{Kauch2019arXiv}. One crucial point is that vertex corrections become less relevant for high frequencies, because complicated diagrams disappear due to the decay of the Green function $G \sim 1/\omega_n$ in high frequencies \cite{Wentzell2020}. Eventually the asymptotic value of the diagonal structure becomes the same as the mean-field-like value. In this sense, with fixed spin fluctuation instability, the vertex corrections suppress the intensity of the pairing vertex at low frequencies. Such a screening decreases the transition temperature by orders of magnitude since the low-frequency value is the most important for Fermi surface instabilities, such as superconductivity. These results suggest the importance of the vertex structure of the pairing interaction for the quantitative estimation of the superconducting transition temperatures.


Considering the dynamical screening effect, we can now compare computational results with experiments quantitatively. Fig.~5 shows the superconductivity phase diagram of the (infinite-layer) nickelate superconductor Sr$_{0.2}$Nd$_{0.8}$NiO$_2$ \cite{Kitatani2020}. The theoretical (D$\Gamma$A) result (yellow circles) is roughly consistent with the experimental value of 15K at 20\%-doping (red diamond) \cite{Li2019}. Furthermore, the $T_c$ dome structure around 20\%-doping is also compatible with subsequent experiments \cite{Li2020,Zeng2020,Lee2022arxiv} (please see the review article \cite{Held2022} for details of nickelate calculations). We note that the ignored $d$-wave pairing fluctuation effect and the interlayer coupling will somewhat decrease $T_c$ while we expect these are secondary (minor) effects as mentioned in Sec.~\ref{sec:MW}. The D$\Gamma$A result was also applied to analyze the recently found quintuple-layer nickelate superconductivity \cite{Pan2021,Worm2021}.

\begin{figure}[t!]
        \centering
                \includegraphics[width=0.6\linewidth,angle=0]{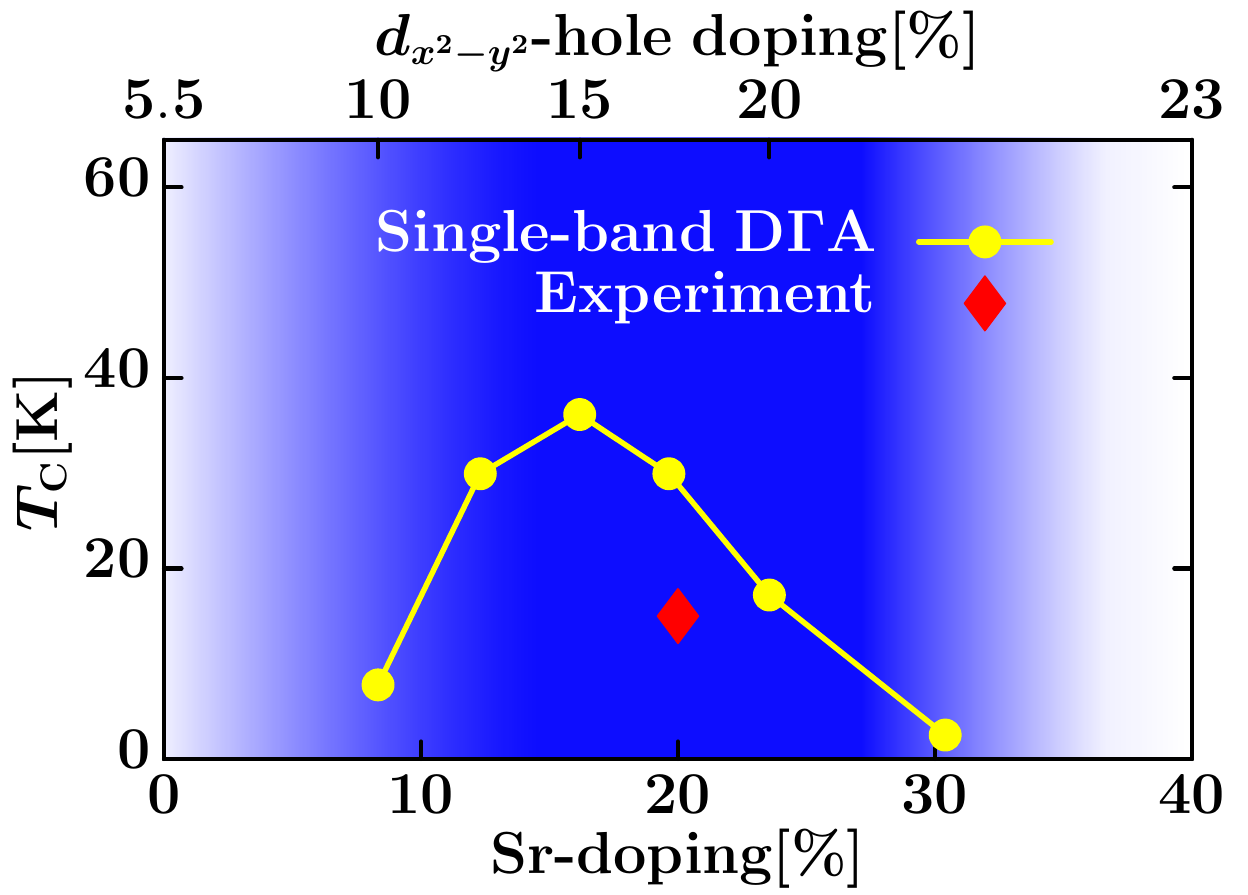}
        \caption{The phase diagram ($T_{\rm c}$ vs. doping) for the nickelate superconductor  Sr$_{0.2}$Nd$_{0.8}$NiO$_2$. Theoretical calculation results for U=2D (yellow line) and the experimental result (red diamond) are shown. The upper axis corresponds to the doping level for the most relevant $d_{x^2-y^2}$ band estimated from the multi-orbital DFT+DMFT study. Taken from \cite{Kitatani2020}.}
        \label{fig:nickelate}
\end{figure}

The dynamical vertex structure \cite{Kitatani2019} may also change the qualitative nature of the pairing. Usually, the instability odd-frequency pairing is weak because of the node at $\omega=0$. On the other hand, there is a peak at $\omega=0$ for the even-frequency pairing, which is favorable for a Fermi surface instability. Since the suppression of $T_c$ by the dynamical vertex structure occurs in low frequencies, it would more strongly affect the even-frequency pairing, reducing the mentioned advantage of the even-frequency pairing and relatively supporting the odd-frequency pairing. This point is further confirmed in Ref.~\cite{Kauch2019arXiv}, where the author analyzed the dynamical screening effect on  both even/odd-frequency-pairing explicitly. Indeed, the odd-frequency superconductivity is observed with the dual-fermion approach for the Kondo-lattice model \cite{Otsuki2015}. There, a similar mechanism may enter while the model and relevant fluctuation are pretty different. Regarding the dynamical structure of the vertex, it would also be an interesting question about the relation between this dynamical screening effect and studies of non (simple-)bosonic pairing glue indicated from the real frequency structures \cite{Sakai2016}.

\section{Summary and outlook}
In this article, we reviewed research for superconductivity obtained by diagrammatic extensions of the dynamical mean-field theory. These approaches can simultaneously capture the effects of strong correlations and spatial fluctuations, which are both relevant for describing layered or thin-film superconductivity. As a result, the dome structure of $T_{\rm c}$ observed in experiments can be described. It is also suggested that we are now able to capture the phase diagram of unconventional superconductivity, not only qualitatively but even quantitatively. An in-depth analysis shows that the dynamical screening of the pairing vertex reduce the superconducting transition temperature by one order of magnitude. We can also apply these approaches to explore the possibility of qualitatively exotic states, e.g., odd-frequency pairing.

In the future, further analyses of the interrelationship among different non-local fluctuations are desirable. Indeed, it has been gradually recognized that the charge (stripe) instability is also relevant for cuprate superconductors both experimentally and theoretically. Stripe instabilities may even dominate the superconductivity in some cases \cite{Darmawan2018,Ohgoe2020,Qin2020}. 
Diagrammatically, spin-fluctuation-assisted charge instability has been proposed and studied recently but rarely starting from DMFT \cite{Fleck2000,Otsuki2014,Kauch2020}. In such systems where many fluctuations compete, taking all channels and their mutual coupling into account will become more important and is possible through the parquet equations. Another direction is the extension to multi-orbital systems to study realistic systems and explore new physics, e.g., orbital fluctuation/Hund physics. Both of these directions are performed already in high-temperature regimes \cite{Kaufmann2021,Kauch2020,Galler2016,Pickem2020}. However, they have not been applied to low-temperature phenomena like unconventional superconductivity. For extending such studies  to the  low-temperature regime, numerical advance is crucial (c.f. Appendix).

\appendix 

\section{Detail formulations of high-frequencies contributions}
One big problem for treating the vertex function is convergence against the number of Matsubara frequencies. For example, we usually use a few thousand points when calculating the unconventional superconductivity by using FLEX. On the other hand, we can only handle a few hundred points for the vertex. Here, we will first explain the details of the schemes used in Ref.~\cite{Kitatani2019} (which we just briefly explained in the Supplemental Section S.II of that paper). This is a simpler version of the pioneering work \cite{Kunes2011}, but instead, can be easily extended toward beyond DMFT and seems to be enough even for describing unconventional superconductivity. Afterwards, we will discuss our results and the relations to other recent progresses.

Similar to \cite{Kunes2011}, we separate the frequency range into the directly calculated range and the region supplementing asymptotic structures to accelerate the convergence against frequency box size ($\nu,\nu',\omega$)-range. On top of the exactly treated region ($n_{\rm core}$), we consider the contribution from the bare $U$ of the irreducible vertex $\Gamma$ for large size ($n_{\rm outer}$) box and further consider the  bare bubble contribution for the susceptibility within an even larger $n_{\rm asympt}$ range. The advantage of this kind of procedure is that we only need to solve $n_{\rm outer}$ size of the local inverse Bethe-Salpeter equation once, and other parts require only $n_{\rm core}$ size calculations. Therefore we can increase $n_{\rm outer}$ up to more than 1000 points, and there, the Green's function is smoothly connected to the asymptotic behavior $1/(i\omega_{\rm n})$ (i.e., we can increase $n_{\rm asympt} \rightarrow \infty$).

Later, we will focus on the Bethe-Salpeter equation in one specific channel, which is enough for treating ladder extensions. Then, the equations are independent for each bosonic frequency $\omega$, and we can assume the vertex function as the matrix. Here, we first make some rules for changing the size of the matrix. For a large matrix $A$, we define $A_{\rm x}, ({\rm x}={\rm core}, {\rm outer}, {\rm asympt})$ as a smaller size ($n_{\rm X}$) matrix as,
\begin{align}
	[X_{\rm x}]_{ij} = X_{ij} (-n_{\rm x} \le i,j \le n_{\rm x}),
\end{align}
and when we sum up matrices of different size, we supplement $0$ for mismatching regions. One remark is that ``the size change and matrix inverse do not commute with each other $A_{\rm core}^{-1} \equiv (A_{\rm core})^{-1} \neq (A^{-1})_{\rm core}$, and we can change these order if the matrix is block diagonalized". This is an essential point regarding the frequency box size convergence, as we will see in the following.



\subsection{Local part}
\begin{figure}[t!]
        \centering
             \includegraphics[width=0.8\linewidth,angle=0]{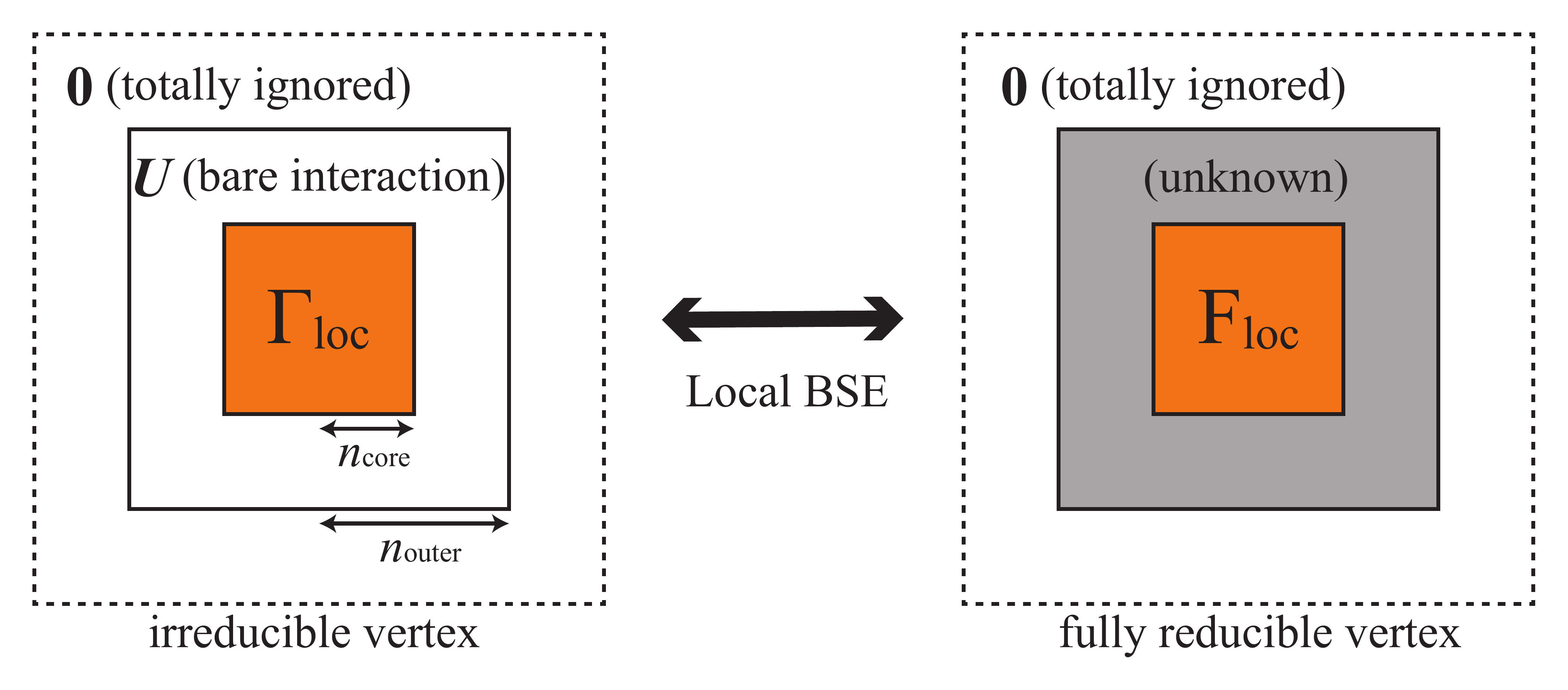}
        \caption{Schematic picture of how to extract the irreducible vertex. With fixed bosonic frequency, the vertex depends on two Matsubara frequencies, and we map these dependencies to a plane. Here, we need to solve the inverse Bethe-Salpeter equation from the fully reducible vertex $F_{\rm loc}$ (which is obtained by the impurity solver). Solid lines indicate the frequency cutoff, which is necessary to perform practical calculations. The standard way takes the orange region ($n_{\rm core}$-Matsubara points) only and ignores outsides. Our approach takes some buffer region where we only consider the bare interaction as irreducible vertex and determine $\Gamma_{\rm loc}$ from the Bethe-Salpeter equation with $n_{\rm outer}$-Matsubara points. Nevertheless, we can extract $\Gamma$ deterministically (without a  self-consistent calculation) as written in the text.}
        \label{fig:local-BSE}
\end{figure}

First, we need to extract the local $\Gamma^{\rm loc}_{\rm core}$ from the generalized susceptibility of the impurity solver  $\chi^{\rm loc}_{\rm solver}$ after convergence of the DMFT calculation. They are related by the local Bethe-Salpeter equation as
\begin{equation}
(\chi^{\rm loc})^{-1} = (\chi^{\rm loc}_0)^{-1} + \Gamma^{\rm loc}.
\label{eq:BSE-normal}
\end{equation}
Now, we want to consider the outside bare $U$ contribution for $\Gamma_{\rm loc}$. Instead of simply solving the $n_{\rm core}$-size Bethe-Salpeter equation, we need to solve the $n_{\rm outer}$-size Bethe-Salpeter equation as
\begin{equation}
(\chi^{\rm loc}_{\rm outer})^{-1} = (\chi^{\rm loc}_{0,{\rm outer}})^{-1} + ({\cal U}_{\rm outer} + \delta\Gamma^{\rm loc}_{\rm core}),
\label{eq:BSE-withbareU}
\end{equation}
where ${\cal U}$ is the matrix representation of the bare interaction and $\delta\Gamma$ is the irreducible vertex after subtracting the bare-$U$ contribution. $\chi_{\rm outer}$ must be the same as $\chi^{\rm loc}_{\rm solver}$ for the core region, while we do not know outsides as shown in Figure~\ref{fig:local-BSE}.
Here, we can ignore the bare bubble contribution outside $n_{\rm outer}$, since $\chi$ is diagonalized and doesn't affect $\Gamma_{\rm core}$ (i.e., $(\chi^{\rm loc}_{\rm asympt})^{-1}_{\rm outer}=(\chi^{\rm loc}_{\rm outer})^{-1}$).

At a glance, this inverse BSE is not well defined due to unknown regions. Since we fix the outer region of the irreducible vertex as the bare interaction, we can obtain $\Gamma^{\rm loc}_{\rm core}$ by one-shot calculation as follows. 
First, we 
define ${\tilde \chi}^{\rm loc}_{\rm outer}$ as
\begin{equation}
({\tilde \chi}^{\rm loc}_{\rm outer})^{-1} = (\chi^{\rm loc}_{0, {\rm outer}})^{-1} + {\cal U}_{\rm outer}.
\label{eq:def-chitilde}
\end{equation}
We consider ${\tilde \chi}^{\rm loc}$  for the full $n_{\rm outer}$, because we do not use $\chi^{\rm loc}_{\rm solver}$ here. From Eqs.~(\ref{eq:BSE-withbareU}) and (\ref{eq:def-chitilde}), the currently considered BSE is now represented as,
\begin{equation}
    \delta \Gamma^{\rm loc}_{\rm core}
    =
    (\chi^{\rm loc}_{\rm outer})^{-1}_{\rm core}-(\tilde{\chi}^{\rm loc}_{\rm outer})^{-1}_{\rm core}.
\label{eq:localBSE-nouter}
\end{equation}
From here on, we need to construct the equation only using $n_{\rm core}$ size.
Focusing on the $n_{\rm core}$ range for $(i,j \in {\rm core})$, we obtain
\begin{align}
[\chi^{\rm loc}_{\rm outer}]_{ij} &= [{\tilde \chi}^{\rm loc}_{\rm outer}]_{ij} 
- \sum_{m,n \in {\rm outer}} [{\tilde \chi}^{\rm loc}_{\rm outer}]_{im} [\delta \Gamma^{\rm loc}_{\rm core}]_{mn} [\chi^{\rm loc}_{\rm outer}]_{nj},
\label{eq:local-nouter-size2}\\
&= [{\tilde \chi}^{\rm loc}_{\rm outer}]_{ij} 
- \sum_{m,n \in {\rm core}} [{\tilde \chi}^{\rm loc}_{\rm outer}]_{im}
[\delta \Gamma^{\rm loc}_{\rm core}]_{mn} [\chi^{\rm loc}_{\rm outer}]_{nj}.
\label{eq:local-ncore-size2}
\end{align}
From Eq.~(\ref{eq:local-nouter-size2}) to Eq.~(\ref{eq:local-ncore-size2}), 
we use the fact that $\delta \Gamma^{\rm loc}_{\rm core}$ have finite value only for the $n_{\rm core}$ range. Then, we finally obtain the $n_{\rm core}$-size equation as $(\chi^{\rm loc}_{\rm outer})_{\rm core}$ should be equal to $\chi^{\rm loc}_{\rm solver}$ which is only defined on $n_{\rm core}$ 
\begin{align}
\delta \Gamma^{\rm loc}_{\rm core} = (\chi^{\rm loc}_{\rm solver})^{-1} -
								   (({\tilde \chi}^{\rm loc}_{\rm outer})_{\rm core})^{-1}.
\label{eq:localBSE-ncore}
\end{align}
Once we obtain $\delta \Gamma^{\rm loc}_{\rm core}$, we can get $\chi_{\rm outer}$ by the Bethe-Salpeter equation. While they look similar for Eqs.~(\ref{eq:localBSE-nouter}) and (\ref{eq:localBSE-ncore}), please note that  each term is different. 

Indeed, the difference of the first terms of Eq.~(\ref{eq:localBSE-nouter}) and Eq.~(\ref{eq:localBSE-ncore}) [i.e., 
$(\chi^{\rm loc}_{\rm outer})^{-1}_{\rm core} \neq (\chi^{\rm loc}_{\rm solver})^{-1}$] 
directly indicates the box-size convergence problem discussed here (i.e., the central structure of the irreducible vertex depends on how many Matsubara grids are used for the inverse calculation). In this regards, naive expectation of  Eq.~(\ref{eq:localBSE-ncore}) is that $\delta \Gamma$ would be more stable against the frequency-size, since $\tilde{\chi}$ has somewhat similar box-size dependence as $\chi$ and they will be cancelled out (instead subtracting box-size independent $\chi_0^{-1}$ for obtaining $\Gamma$ in the usual protocol).

\begin{figure}[t!]
        \centering
                \includegraphics[width=1.0\linewidth,angle=0]{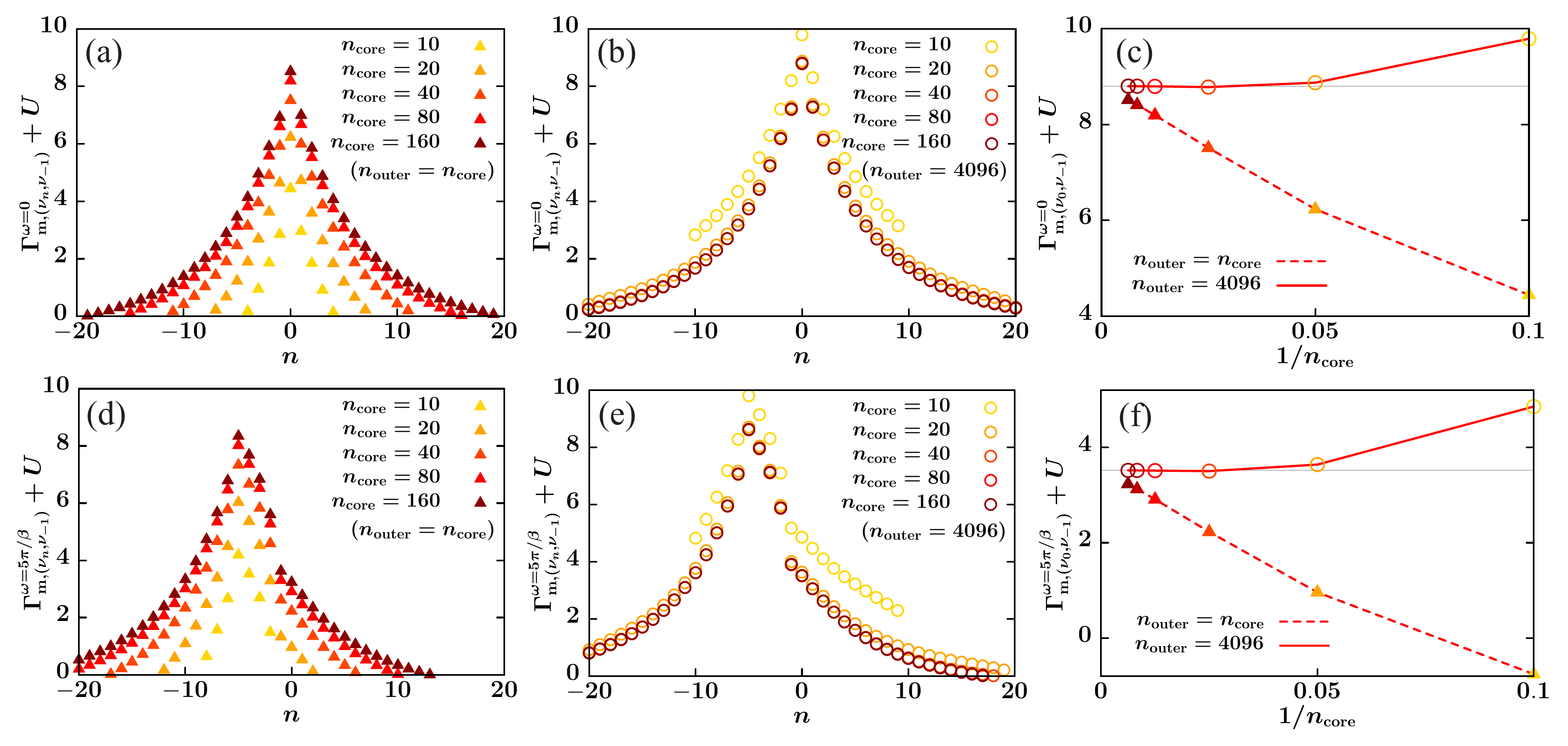}
        \caption{Matsubara frequency dependence of the irreducible vertex in the magnetic channel $\Gamma_{\rm m}$ for (a,b,c) $\omega=0$ and (d,e,f) $\omega=5\pi/\beta$ with various cutoff frequencies $n_{\rm core}$ using (a,d) the standard way and (b,e) the described procedure  with a buffer region ($n_{\rm outer}=4096$). (c,f) $n_{\rm core}$ dependence of the result for $n=0$ results in (a-d).}
        \label{fig:gammaconv}
\end{figure}

In Figure \ref{fig:gammaconv}, we checked the current procedure for the DMFT result on the square lattice Hubbard model at $U=1.5D,n=0.85,\beta D=100$. We compare particular cuts of the irreducible vertex in the magnetic channel: $\Gamma_m$, for (a,d) simply solving the inverse Bethe-Salpeter equation (without supplementing bare $U$) and (b,e) with supplementing bare-$U$ contribution as Figure \ref{fig:local-BSE}. When simply solving the inverse Bethe-Salpeter equation, the result is not well converged even for $n_{\rm core}=160$. On the other hand, the results in (b,e) are well converged already around $n_{\rm core}=40$. In Figure (c,f), we further check the frequency range dependence of $n=0$ (lowest Matsubara) values for (a-d). We can see that convergence is pretty good when we supplement the bare-$U$ contribution. The converged value is indeed almost the same as the extrapolated value from the result of simply solving the inverse Bethe-Salpeter equation.

\subsection{Non-local part}
Let us move on to the formulation of the nonlocal diagrammatic extension. For evaluating the superconductivity eigenvalues, we need to calculate the nonlocal self-energy $\Sigma$ and the ladder expanded vertex F (for obtaining the pairing interaction $\Gamma_{\rm pp}$ from $\Gamma_{\rm pp} \equiv F -\Phi^{\rm DMFT}_{\rm pp}$ where $\Phi^{\rm DMFT}_{\rm pp}$ is the reducible part of the two-particle vertex) given as \cite{Rohringer2018,Kitatani2019}
\begin{align}
\Sigma(k) &= \Sigma_{\rm Hartree}+ \frac{U}{2\beta^2}\sum_{q} 
[\gamma_{\rm d}(k,q)-3\gamma_{\rm m}(k,q)+U\gamma_{\rm d}(k,q)\chi_{\rm d}(q)
+3U\gamma_{\rm m}(k,q)\chi_{\rm m}(q) +2 \nonumber\\
&-\sum_{\nu'} (F^{\rm loc}_d-F^{\rm loc}_m) G(k^{\prime})G(k^{\prime}+q)]G(k+q), \\
F^{\nu,\nu^{\prime}}_{\rm r} &= (\chi^{\nu}_0)^{-1}
    \left[ 
    \delta_{\nu,\nu^{\prime}} -\chi^{*,\nu\nu^{\prime}}_{\rm r} (\chi^{\nu^{\prime}}_0)^{-1}
    \right] 
    +U_{\rm r}(1-U_{\rm r}\chi_{\rm r})\gamma_{\rm r}^\nu \gamma_{\rm r}^{\nu^{\prime}}
\end{align}
where $\chi_{\rm r}(q),({\rm r}={\rm d},{\rm m})$ is the physical susceptibility defined as $\chi_{\rm r}^{\rm phys}(q) = \sum_{\nu,\nu'}[\chi_0^{-1}+\Gamma_{\rm r}]^{-1}$, $\chi_{\rm r}^{*} \equiv [\chi_{\rm r}^{-1}-{\cal U}_{\rm r}]^{-1}$, $\gamma_{\rm r}^{\nu} \equiv (\chi_0)^{-1} \sum_{\nu'} \chi^{*,\nu\nu'}_{\rm r}, U_{\rm d/m} \equiv \pm U$.
Since we want to consider high-frequency bare-$U$ contributions for $\Gamma^{\rm loc}$,
we write this effect explicitly like $\chi_{\rm X},\gamma_{\rm X},\chi^{*}_{\rm X}$
and then, we explain how to calculate these quantities with considering asymptotic behavior
with small ($n_{\rm core}$) size vertex calculations. Hereafter, we omit some subscripts ($\omega,{\rm d,m}$), since they are independent in the particle-hole channel and the following calculations apply equally for all degree of freedom.

We first define $\chi_X$ and $\chi_X^{*}, (X \in {\rm core, outer, asympt})$ as
\begin{align}
\chi^{-1}_X  &= \chi^{-1}_{0,X} + \Gamma_X, \\
(\chi^{*})^{-1}_X &= \chi^{-1}_X - {\cal U}_X. \label{eq:chistar}
\end{align}
These equations must be satisfied for each size $X$, i.e.,  for $n_{\rm core}, n_{\rm outer}$, and $n_{\rm asympt}$, respectively.
From Eq.~(\ref{eq:chistar}), we can derive relations like
\begin{align}
\chi_X^{*}{\cal U} &= \frac{\chi_X {\cal U}}{1 - U_{\rm r}{\overline {\chi_X}}} \\
\chi_X^{*} &= \chi_X +  \frac{\chi_X {\cal U}\chi_X}{1 - U_{\rm r}{\overline {\chi_X}}},
\end{align}
where the overline means the summation like 
$\overline{\chi_{\nu,\nu^{\prime}}} \equiv \sum_{\nu,\nu^{\prime}} \chi_{\nu,\nu^{\prime}}$.
Different size vertices are related as
$(\chi_{\rm asympt})_{\rm outer}= \chi_{\rm outer}, (\chi^{*}_{{\rm outer}})_{\rm core}= \chi^{*}_{\rm core}$, while $(\chi_{\rm outer})_{\rm core} \neq \chi_{\rm core}, (\chi^{*}_{\rm asympt})_{\rm outer} \neq \chi^{*}_{\rm outer}$.

Then, the necessary quantities are calculated from the core-size diagrammatic expansion ($\chi^*_{\rm coce}$) as follows.
\begin{align}
\chi_{\rm phys} &= {\overline {\chi_{\rm asympt}}}, \nonumber \\
	&= {\overline {\chi_{\rm outer}}} + {\overline {\chi_{0,{\rm asympt}}}}-{\overline {\chi_{0,{\rm outer}}}}, \nonumber \\
	&= (({\overline {\chi^{*}_{\rm outer}}})^{-1}+U_{\rm r})^{-1} + {\overline {\chi_{0,{\rm asympt}}}}-{\overline {\chi_{0,{\rm outer}}}}, \nonumber \\
	&= (({\overline {\chi^{*}_{\rm core}}} + {\overline {\chi_{0,{\rm outer}}}} - {\overline {\chi_{0,{\rm core}}}})^{-1}+U_{\rm r})^{-1} 
			+ {\overline {\chi_{0,{\rm asympt}}}}-{\overline {\chi_{0,{\rm outer}}}}.
\end{align}

\begin{align}
[U_{\rm r}\gamma_{\rm asympt}]_{\rm core}
	=& [\chi_0^{-1} \chi^{*}_{\rm asympt}{\cal U}], \nonumber \\
	=& \frac{[\chi_0^{-1} \chi_{\rm asympt}{\cal U}]_{\rm core}}{1-U_{\rm r}{\overline {\chi_{\rm asympt}}}}, \nonumber \\
	=& \frac{[\chi_0^{-1} \chi_{\rm outer}{\cal U}]_{\rm core}}{1-U_{\rm r}{\overline {\chi_{\rm asympt}}}}, \nonumber \\
	=& \chi_0^{-1}[\chi^{*}_{\rm outer}]_{\rm core}{\cal U} \frac{1-U_{\rm r}{\overline {\chi_{\rm outer}}}}{1-U_{\rm r}{\overline {\chi_{\rm asympt}}}}, \nonumber \\
	=& \chi_0^{-1} \chi^{*}_{\rm core} {\cal U} \frac{1-U_{\rm r}{\overline {\chi_{\rm outer}}}}{1-U_{\rm r}{\overline {\chi_{\rm asympt}}}}, \\
[U_{\rm r}\gamma_{\rm asympt}]\; (\nu_{n_{\rm core}} <\nu< \nu_{n_{\rm outer}})  
	=& U_{\rm r} \frac{1-U{\overline {\chi_{\rm outer}}}}{1-U_{\rm r}{\overline {\chi_{\rm asympt}}}}.
	\label{eq:gamma_asympt}
\end{align}

\begin{align}
[\chi^{*}_{\rm asympt}]_{\rm core} 
	=& [ \chi_{\rm asympt} + \frac{\chi_{\rm asympt}{\cal U}\chi_{\rm asympt}}{1-U_{\rm r}{\overline {\chi_{\rm asympt}}}} ]_{\rm core}, \nonumber \\
	=& [ \chi_{\rm outer} +  \frac{\chi_{\rm asympt}{\cal U}\chi_{\rm asympt}}{1-U_{\rm r}{\overline {\chi_{\rm asympt}}}} ]_{\rm core}, \nonumber \\
	=& \chi^{*}_{\rm core} - \chi^{*}_{\rm core}{\cal U}\chi^{*}_{\rm core}(1-U_{\rm r}{\overline {\chi_{\rm outer}}})
							+\chi^{*}_{\rm core}{\cal U}\chi^{*}_{\rm core} \frac{(1-U_{\rm r}{\overline {\chi_{\rm outer}}})^2}{(1-U_{\rm r}{\overline {\chi_{\rm asympt}}})}.
\end{align}
Eq.~(\ref{eq:gamma_asympt}) does not obey the well known asymptotic behavior $\gamma^{\nu} \rightarrow 1$ as $\nu \rightarrow \infty$ \cite{Katanin2009}. This is because we consider the infinite range of bare-$U$ interaction in Eq.~(\ref{eq:chistar}) while we take $n_{\rm outer}$ range for the whole formalism. Nevertheless, there is no ambiguity and this just comes from how to express the same quantity. Please note that $\gamma$ appears as $\gamma(1-U_{\rm r}\overline{\chi})$ in the self-energy calculation and $\gamma_{\rm asympt}(1-U_{\rm r}\overline{\chi_{\rm asympt}})=\gamma_{\rm outer}(1-U_{\rm r}\overline{\chi_{\rm outer}}),\gamma_{\rm outer} \rightarrow 1$. Then the current expression is fully consistent even if considering up to $n_{\rm outer}$ range, and the exact asymptotic behavior holds: $\gamma_{\rm outer} \rightarrow 1$. On the other hand, we expect that $\chi_{\rm asympt}$ converges slightly faster, and the current expression is better for considering the physical susceptibility itself (for example, when applying sum-rule for them \cite{Katanin2009}).

\begin{figure}[hbtp]
        \centering
                \includegraphics[width=1.0\linewidth,angle=0]{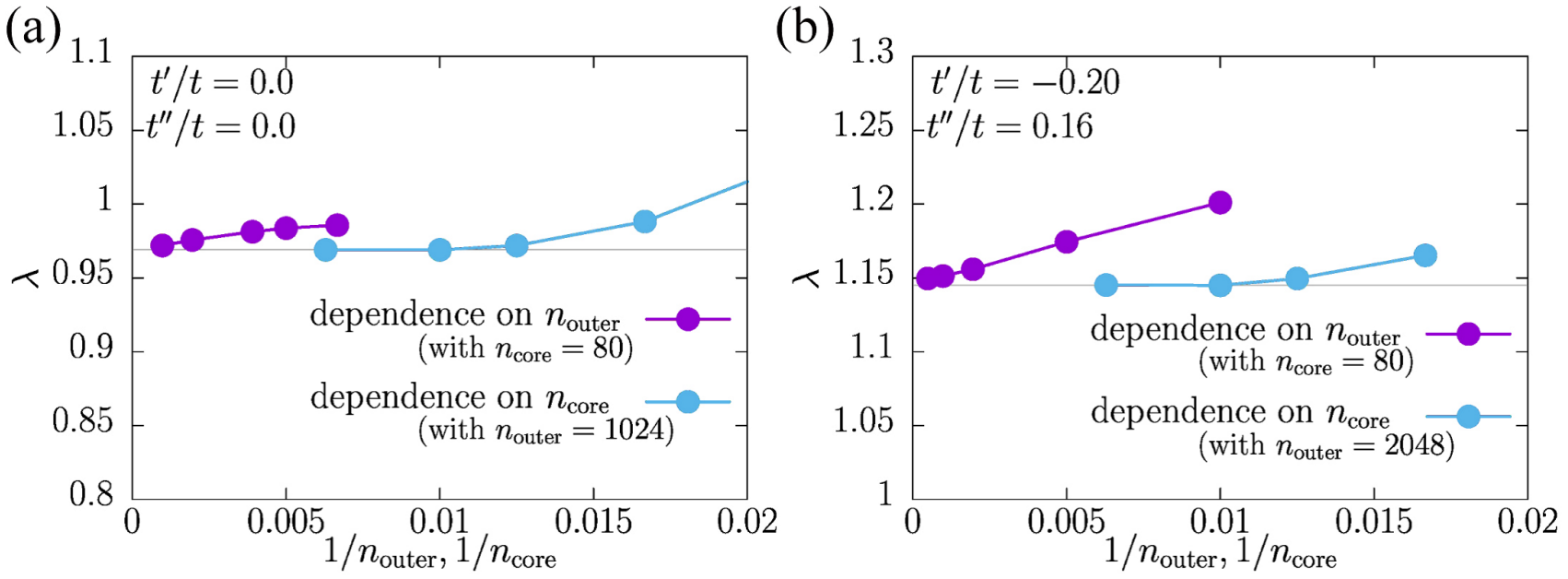}
        \caption{Matsubara frequency cutoff ($n_{\rm core}, n_{\rm outer}$) dependence of leading superconductivity eigenvalue in the square lattice Hubbard model for (a) $n=0.825, t^{\prime}/t=t^{\prime\prime}/t=0$ and (b) $n=0.90,t^{\prime}/t=-0.20,t^{\prime\prime}/t=0.16$. The purple lines show $n_{\rm outer}$-dependence with fixed $n_{\rm core}=80$ and the blue lines show $n_{\rm core}$-dependence with fixed $n_{\rm outer}=1024$ for (a) and $n_{\rm outer}=2048$ for (b). Horizontal lines are guides to the eyes for the almost converged result at $n_{\rm core}=159$. Taken from Ref.~\cite{Kitatani2019}}
        \label{fig:lambdaconv}
\end{figure}

Finally, let's look at how the current formalism works for evaluating unconventional superconductivity. We show the numerical result employing the current scheme in Figure \ref{fig:lambdaconv} for $\beta D=400$ \cite{Kitatani2019}. Please note that it is impossible to get well converged results with simple implementation for such low temperatures. We can see that we can increase $n_{\rm outer}$ around several thousand, and, thus, we only need a few hundred Matsubara grids for the direct treatment of the vertex.

\subsection{Discussion}
Figures \ref{fig:gammaconv} and \ref{fig:lambdaconv} indicate that such the simple framework schematically vizualied in Fig.~\ref{fig:local-BSE} efficiently accelerates the convergence. Indeed, figure \ref{fig:lambdaconv} shows that the size of bare-U itself is much more important than the rest of the detailed asymptotics for capturing the high-frequency contributions. This fact can be understood as follows: For high frequencies, the fermion-boson three points vertex is reduced to the bare interaction, and in such regions, simply the summation of the vertex is an important quantity, for example, a part of the ladder: $\gamma \chi_0 \Gamma \chi_0 \gamma$ (c.f., \cite{Stepanov2016,Stepanov2019}) becomes ${\cal U}\chi_0 \Gamma \chi_0 {\cal U}=U^2 \sum_{\nu,\nu^{\prime}} \chi_0^{\nu} \Gamma^{\nu,\nu^{\prime}} \chi_0^{\nu^{\prime}}$ as $\gamma \rightarrow {\cal U}$. 
For irreducible vertices, many asymptotic structures persist but at a specific frequency. In the most case, the bare-interaction contribution dominates if we sum up against Matsubara frequencies except that the fluctuation in another channel diverges. Therefore, we expect supplementing the bare interaction for the irreducible vertex works efficiently at least for the most relevant channel.

There are many other works for treating the complicated vertex more efficiently. In addition to the pioneering work \cite{Kunes2011}, Ref.~\cite{Li2016} employed the vertex at large frequency to extrapolate it to its asymptotics in parquet calculations,  Ref.~\cite{Tagliavini2018} studied the effect of supplementing the vertex asymptotics \cite{Wentzell2020} for all channels at the local level. Ref.~\cite{Katanin2020} tries to use the analytical expression to take the large size limit of the outer-box. 
There has been some progress recently \cite{Wallerberger2021}
solving the Bethe-Salpeter equation in the intermediate representation \cite{Shinaoka2017,Shinaoka2020}. Also a compactification of the vertex in momentum space has been suggested using a form factor expansion \cite{Eckhardt2020}. Other than these, many kinds of the vertex decomposition have been proposed \cite{Otsuki2019,Mizuno2021,Mizuno2022}. 
Ref.~\cite{Krien2020,Krien2021} extended the current (supplementing bare-$U$) scheme to the parquet equations by using single-boson-exchange (SBE) decomposition \cite{Krien2019} based on a concept of the irreducibility against the bare interaction \cite{Katanin2009}. Please note that SBE graphically has a similar structure as Eq.~(\ref{eq:lowrankF}), and clarifying the relation between them would be important. Indeed, for the matrix, eigenvalue decomposition (more generally the singular value decomposition) is straightforward and almost equivalent to the solving the Bethe-Salpeter equation (c.f., Eqs.~(\ref{eq:lowrankGamma},\ref{eq:lowrankF})), while it is not trivial for the relation between solving parquet equations and higher rank tensor decompositions of the irreducible vertices. 
A big step is still an efficient implementation for the parquet equations where memory constraints become even more relevant than for the ladder calculations and  which likely, requires some further physical insight.

\section*{Acknowledgments}
We thank Alessandro Toschi, Hideo Aoki, Agnese Tagliavini, Friedrich Krien, Anna Kauch, Paul Worm and Liang Si for illuminating discussions. We thank Elio K{\"o}nig for critically reading the manuscript. We acknowledge the financial support by Grant-in-Aids for Scientific Research (JSPS KAKENHI) [Grant No. 19H05825, JP20K22342 and JP21K13887], and the Austrian Science Funds (FWF) through project P~32044.

\section*{References}
\bibliographystyle{unsrt}
\bibliography{main}

\end{document}